\documentclass[10pt, twocolumn, unsortedaddress, superscriptaddress, raggedfooter, showpacs, prb]{revtex4}
\usepackage{amsmath}
\usepackage{graphicx}
\begin{document}

\title{On the Coexistence Magnetism/Superconductivity in the Heavy-Fermion Superconductor CePt$_3$Si}

\author{A. Amato}
\email{alex.amato@psi.ch}
\affiliation{Laboratory for Muon-Spin Spectroscopy, Paul Scherrer Institute, CH-5232 Villigen PSI, Switzerland}
\author{E. Bauer}
\affiliation{Institut f\"ur Festk\"orperphysik, Technische Universit\"at Wien, A-1040 Wien, Austria}
\author{C. Baines}
\affiliation{Laboratory for Muon-Spin Spectroscopy, Paul Scherrer Institute, CH-5232 Villigen PSI, Switzerland}
\date{\today}
\begin{abstract} The interplay between magnetism and superconductivity in the newly discovered heavy-fermion superconductor CePt$_3$Si has been investigated using the zero-field $\mu$SR technique. The $\mu$SR data indicate that the whole muon ensemble senses spontaneous internal fields in the magnetic phase, demonstrating that magnetism occurs in the whole sample volume. This points to a microscopic coexistence between magnetism and  heavy-fermion superconductivity. 
\end{abstract}

\maketitle
\section{Introduction}
In recent years, strongly correlated electron systems have played a leading role in solid state physics. The importance of this research field is illustrated by the discovery of novel phases in metals, intermetallics and oxides at low temperatures. One of the most relevant example is the discovery of unconventional superconductivity in heavy-fermion systems.

Unconventional superconductivity seems to result from the nature of the mechanism providing the attractive force necessary for the Cooper-pair formation. In conventional superconductors, the electrons are paired in a spin-singlet zero-angular-momentum state ($L=0$), which results from the fact that their binding is described in terms of the emission and absorption of phonons. This leads to the formation of an isotropic superconducting gap in the electronic excitations over the whole Fermi surface. On the other hand, heavy-fermion superconductivity is observed to show a close interplay with magnetic fluctuations. This seems to indicate that the attractive effective interaction between the strongly renormalized heavy quasiparticles in the superconducting heavy-fermion systems is not provided by the electron-phonon interaction as in ordinary superconductors, but rather is mediated by electronic spin fluctuations. This non-conventional (i.e. non-BCS) mechanism is believed to lead to an unconventional configuration of the heavy-fermion superconducting state, which may involve anisotropic, nonzero-angular-momentum states ($L \ne 0$, see Ref.~\onlinecite{sigrist_RMP_1991} 
for a review and references therein). 

An additional feature in a number of systems is the observation of an apparent coexistence of heavy-fermion superconductivity and static magnetism. However, at ambient pressure, such a coexistence was up to recently solely confirmed on uranium-based heavy-fermion systems, and was discarded on cerium-based systems. Such conclusions were deduced from microscopic studies, in particular from the sensitive $\mu$SR technique \cite{amato_RMP_1997}. In this frame, the example of the first discovered heavy-fermion superconductor CeCu$_2$Si$_2$ is examplary as it exhibits a competition between both ground states, i.e. magnetism and superconductivity do not coexist, but appear as two different, mutually exclusive ground states of the same subset of electrons. Such competition  was first discovered by $\mu$SR \cite{luke_PRL_1994,feyerherm_PHYSREVB_1997} and only recently confirmed by neutron studies \cite{stockert_TBP}. 

Recently another Ce-based heavy-fermion system, namely CePt$_3$Si, was found \cite{bauer_PRL_2004} showing antiferromagnetism and superconductivity ($T_{\text N} = 2.2$~K and $T_c = 0.75$~K) at ambient pressure. This material crystallizes in a tetragonal structure (space group $P4mm$) lacking a center of inversion symmetry. This feature is attracting presently much interest since unconventional superconductivity with spin-triplet was to date thought to require such inversion symmetry to obtain the necessary degenerated electron states \cite{anderson_PRB_1984, frigeri_PRL_2004}. In this article, we present $\mu$SR studies aiming to test - at the microscopic level - the coexistence between magnetic and superconducting state. 

\section{Experiment}
CePt$_3$Si was prepared by high frequency melting and subsequently heat treated at 880$^{\circ}$C for one week. Phase purity was evidenced by x-ray diffraction. The $\mu$SR experiments were carried out at the Swiss Muon Source of the Paul Scherrer Institute (Villigen, Switzerland). Measurements were performed on the GPS and LTF instruments of the $\pi$M3 beamline, using a He-flow cryostat (base temperature 1.7~K) and a $^3$He-$^4$He dilution refrigerator (base temperature 0.03~K), respectively. In order to avoid a depolarizing background $\mu$SR signal, the sample was glued onto a high-purity silver holder. Measurements on both instruments were performed on the same sample and showed a very good agreement in the overlapping temperature range. Measurements were performed in zero applied field (ZF), with an external-field compensation of the order of $\pm 20$~mOe.

\section{Results and discussion}

ZF $\mu$SR is a local probe measurement of the magnetic field at the muon stopping site(s) in the sample. If the implanted polarized muons are subject to magnetic interactions, their polarization becomes time dependent, ${\mathbf P}_{\mu}(t)$. By measuring the asymmetric distribution of positrons emitted when the muons decay as a function of time, the time evolution of $P_{\mu}(t)$ can be deduced. The function
$P_{\mu}(t)$ is defined as the projection of ${\mathbf P}_{\mu}(t)$ along the direction of the initial polarization: $P_{\mu}(t) ={\mathbf P}_{\mu}(t)\cdot {\mathbf P}_{\mu}(0)/P_{\mu}(0)=G(t)P_{\mu}(0)$. Hence, the depolarization function $G(t)$ reflects the normalized muon-spin autocorrelation function $G(t)=\langle{\mathbf S}(t)\cdot{\mathbf S}(0)\rangle/S(0)^2$, which depends on the average value, distribution, and time evolution of the internal fields, and therefore contains all the physics of the magnetic interactions of the muon inside the sample \cite{blundell_CONTEMP_1997}. 

\begin{figure}[t]
\includegraphics[width=8cm]{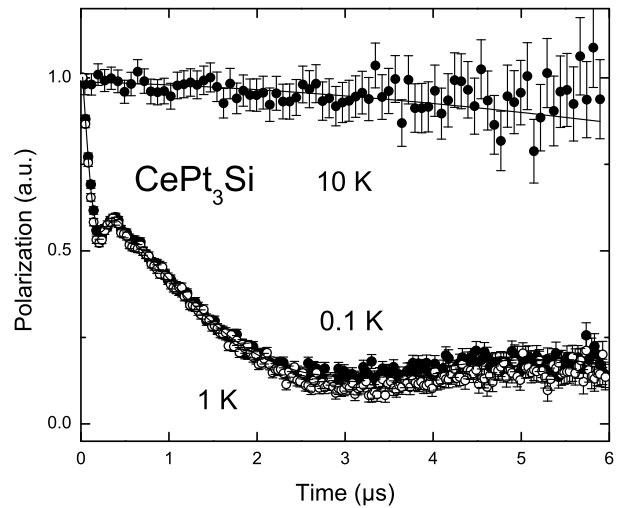}
\caption{\label{figure_cept3si_raw} Example of ZF $\mu$SR signals obtained in polycrystalline CePt$_3$Si in the paramagnetic phase (10~K), the magnetic phase (1~K) and below the superconducting transition (0.1~K). The lines represent fits obtained using Eq.~(\ref{equation_cept3si_kt}) and (\ref{equation_cept3si_afm}). Note that for clarity, the fit for the data obtained at 1~K is ommitted.}
\end{figure}

Above $T_{\text N}$, the time evolution of the muon polarization is best described by the well known Kubo-Toyabe function \cite{kubo_MAGNRES_1967}:  
\begin{equation}
\label{equation_cept3si_kt}
G^{\text {para}}(t) = \frac{1}{3} + \frac{2}{3}(1-\Delta^2 t^2)\exp(-\frac{\Delta^2 t^2}{2})~,
\end{equation}
where $\Delta^2/\gamma_{\mu}^2$ represents the second moment of the local field distribution at the muon site ($\gamma_{\mu}$ is the gyromagnetic ration of the muon). Such a depolarization function is characteristic of a paramagnetic state where the muon depolarization is solely due to the dipolar fields of the nuclear moments ($^{29}$Si and $^{195}$Pt). In the paramagnetic state, the electronic magnetic moments are often not observable by $\mu$SR due to their fast fluctuation rates. Alternatively, the nuclear magnetic moments appear static within the $\mu$SR time window and create a Gaussian field distribution at the muon stopping site, leading to the Kobo-Toyabe depolarization function reported in Eq.~(\ref{equation_cept3si_kt}). Note that this function posses an initial Gaussian character [$\simeq \exp(-\Delta^2t^2)$ for $t \ll \Delta^{-1}$] as observed in the data reported on  Fig.~\ref{figure_cept3si_raw}. Fitting Eq.~(\ref{equation_cept3si_kt}) to the data provides a depolarization rate $\Delta = 0.06~\text{MHz}$ corresponding to field distribution width of $\sim0.7$~G at the muon site, in line with theoretical values computed for several possible stopping sites.  

\begin{figure}[t]
\includegraphics[width=8cm]{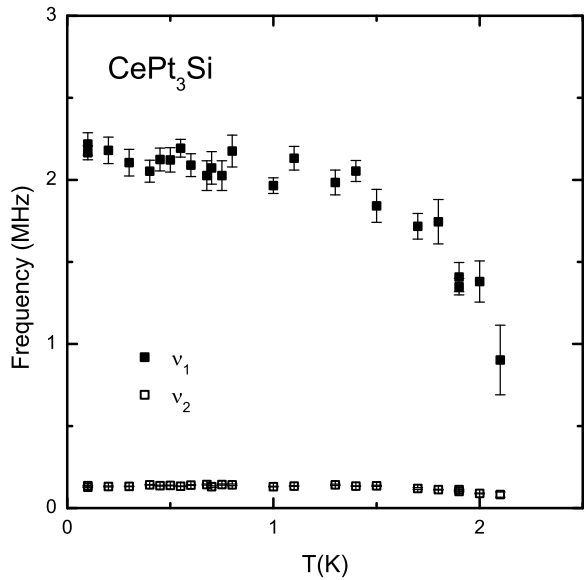}
\caption{\label{figure_cept3si_frequencies} Temperature dependence of the spontaneous $\mu$SR frequencies $\nu_1$ and $\nu_2$ obtained by fitting Eq.~(\ref{equation_cept3si_afm}) to the $\mu$SR data. The measurements were performed in a polycrystalline CePt$_3$Si sample.}
\end{figure}

Below $T_{\text N}$, clear spontaneous oscillations are detected in the $\mu$SR signal indicating the occurrence of static finite magnetic fields sensed by the muons and arising from static electronic magnetic moments. In the antiferromagnetic state, the $\mu$SR signal is best described by the sum of two components, i.e.:
\begin{eqnarray}
\label{equation_cept3si_afm}
G^{\text {AF}}(t) &= &A_1\big[\tfrac{1}{3} \exp(-\lambda_1 t) +\nonumber\\ 
&  & \quad\,\,\,\,\tfrac{2}{3} \exp(-\lambda_1' t) \cos (2 \pi \nu_1 t + \phi_1)\big] +\nonumber\\
&  & A_2\big[\tfrac{1}{3} \exp(-\lambda_2 t) + \nonumber\\ 
&  & \quad\,\,\,\,\tfrac{2}{3} \exp(-\lambda_2' t) \cos (2 \pi \nu_2 t + \phi_2)\big]~.
\end{eqnarray}
These components indicate the presence of two magnetically inequivalent muon stopping sites sensing internal fields $\lvert {\mathbf B}_{\mu}^i\rvert=2\pi\nu_i/\gamma_{\mu}$. As expected for a polycrystalline sample, the ``$1/3$-term'' of each component represents the fraction of the muons possessing an initial polarization along the same direction of the internal field. Therefore, the depolarization rates related to these fractions ($\lambda_i$) reflect solely the internal spins dynamics, whereas the depolarization rates $\lambda_i'$ are ascribed by both dynamical and static effects. The temperature evolution of the spontaneous frequencies $\nu_i$ are reported on Fig.~\ref{figure_cept3si_frequencies}. The values of the frequencies at $T \rightarrow 0$ correspond to internal field values of $\sim$~160~G and 10~G, respectively.

\begin{figure}[b]
\includegraphics[width=7cm]{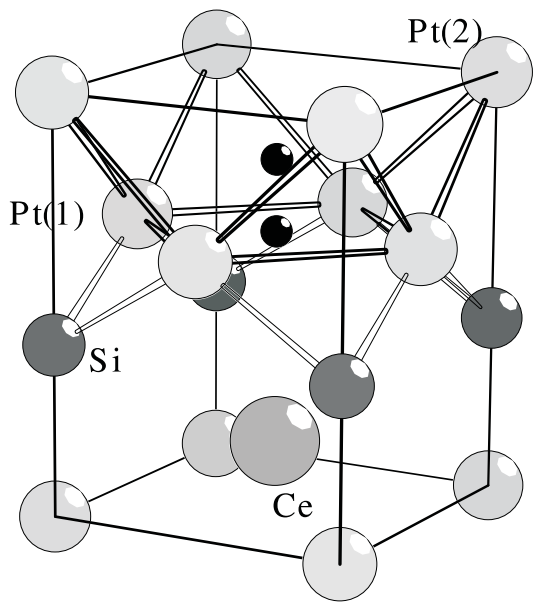} \caption{\label{figure_cept3si_structure} Crystal structure of 
CePt$_3$Si. The smallest spheres represent the muon stopping sites discussed in the text.}
\end{figure}

\begin{figure}[b]
\includegraphics[width=8cm]{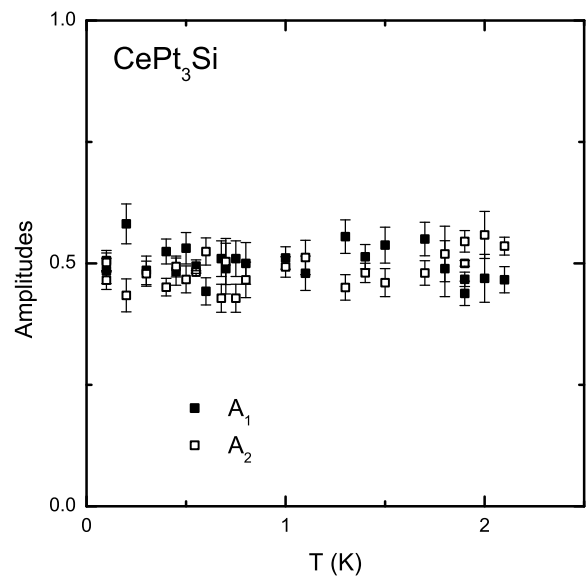} \caption{\label{figure_cept3si_asymmetry} Temperature dependence of the amplitudes $A_1$ and $A_2$ of the spontaneous $\mu$SR frequencies in CePt$_3$Si [see Eq.~(\ref{equation_cept3si_afm})].}
\end{figure}

Very recently, neutron scattering experiments determined the magnetic structure of CePt$_3$Si \cite{metoki_JPCM_2004}. Magnetic Bragg reflections observed at wave vector values ${\mathbf Q} = (0,0,1/2)$ and $(1,0,1/2)$, indicate that magnetic moments align ferromagnetically in the basal plane and stack antiferromagnetically along the $c$ axis with a strongly reduced value of $0.16~\mu_{\text B}$. By considering this magnetic structure, the values of the spontaneous $\mu^+$-frequencies provide information for a tentative determination of the muon stopping sites in the tetragonal structure. By assuming the magnetic moment direction pointing along the $a$ or $b$ axis, the most probable muon sites are located at two different 1(b) Wyckoff-positions, i.e. $(1/2,1/2,0.65)$ for the low-frequency component and $(1/2,1/2,0.82)$ for the high-frequency component. These sites are respectively located in the center of the Pt-plane formed by the Pt(1) ions and between the planes formed by Pt(1) and Pt(2) ions (see Fig.~\ref{figure_cept3si_structure}, notation from Ref.~\onlinecite{bauer_PRL_2004}). In addition, for these sites, the calculated field distributions due to nuclear dipole moments are found in reasonable agreement with the observed depolarization rate $\Delta$ observed in the paramagnetic regime [see Eq.~(\ref{equation_cept3si_kt})]. Note also that both sites have the same multiplicity, which is in line with the observation that $A_1 \simeq A_2$ as shown on Fig.~\ref{figure_cept3si_asymmetry}.

The first relevant observation is that $A_1 + A_2 = 1$ for all temperatures below $T_{\text N}$. This means that the whole muon ensemble is sensing the magnetic state, which in turn unambiguously demonstrates that the \textit{whole} sample volume is involved in the magnetic phase below $T_{\text N}$. Together with thermodynamical studies, demonstrating that superconductivity has a bulk character, the present observation indicates a microscopic coexistence between magnetism and superconductivity. A similar conclusion was very recently drawn from NMR studies performed at different frequencies \cite{yogi_PRL_2004}. Note that the conclusion obtained by $\mu$SR is independent of the exact knowledge of the muon stopping sites. The behavior observed here in CePt$_3$Si is opposite to the one reported for CeCu$_2$Si$_2$ (see above), where the magnetic state is expelled from the sample upon cooling the below $T_c$. The observed coexistence in CePt$_3$Si is reminiscent of the situation observed in U-based heavy-fermion systems, as UPd$_2$Al$_3$ (Ref.~\onlinecite{feyerherm_PRL_1994}) or UNi$_2$Al$_3$ (Ref.~\onlinecite{amato_ZP_1992}), where a model of two independent elctronic subsets, localized and itinerant (responsible for magnetism and superconductivity, respectively), was proposed in view of similar microscopic studies \cite{feyerherm_PRL_1994} and thermodynamics measurements \cite{caspary_PRL_1993}.

\begin{figure}[t]
\includegraphics[width=8cm]{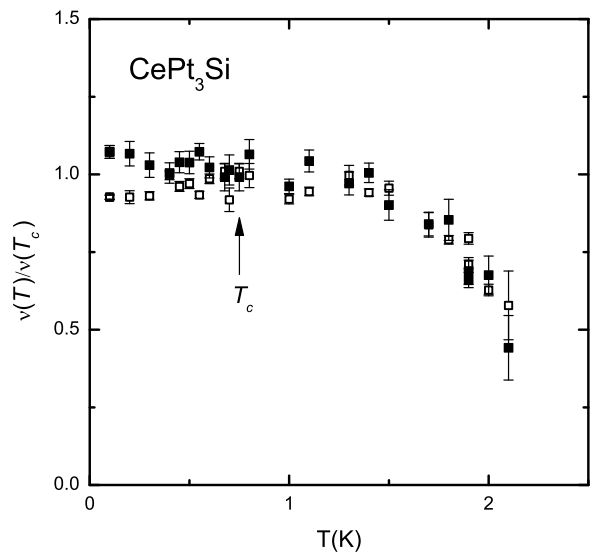} \caption{\label{figure_cept3si_freq_norm} Temperature dependence of the spontaneous $\mu$SR frequencies normalized to their values at $T_c = 0.75$~K. Note the very slight change below $T_c$. The symbols correspond to those of Fig.~\ref{figure_cept3si_frequencies}.}
\end{figure}
Upon cooling the system into the superconducting state, the $\mu$SR data suggest a slight change of the absolute spontaneous internal fields at the muon sites. As shown on Fig.~\ref{figure_cept3si_freq_norm}, one observes, for $T < T_c$, a slight reduction and increase of the low and high frequency signals, respectively. Note that such changes are at the limit of the measurement accuracy. In any case, two possible explanations could be invoked for these changes. The first one would be to consider a coupling between the superconducting and magnetic order parameters, reminiscent to the situation observed in UPt$_3$ \cite{aeppli_PRL_1989}. Alternatively, the frequency changes could have a more simple origin, as the muon is sensing interstitial fields and therefore only indirectly probes the strength of the magnetic order parameter. Hence, in addition to the dipolar interaction, the static 4$f$ magnetic moments will change the spin polarization of the conduction electrons at the muon site \cite{amato_RMP_1997}, which results in an increased hyperfine field action on the muon. Such a contribution is a function of the density of normal electron states and will therefore be affected upon cooling the sample into the superconducting state. Below $T_c$, one expects a decrease in absolute value due to the opening of the superconducting gap. Depending on the muon stopping site and due to the oscillatory character of the RKKY interaction between the static 4$f$ moments and the conduction electrons, a decrease of the hyperfine field contribution can actually lead to either an increase or a decrease of the total internal fields at the muon site, as possibly observed in the present $\mu$SR data. 

\section{Conclusion}
Our zero-field $\mu$SR data have demonstrated the bulk character of the antiferromagnetic state in the heavy-fermion superconductor CePt$_3$Si, suggesting therefore a microscopic coexistence between magnetism and superconductivity. In addition, a slight change of the $\mu$SR response upon cooling the sample below $T_c$ can be ascribed to a coupling of the superconducting and magnetic order parameters and/or to the decrease of the hyperfine contact contribution acting on the muon.
~\\

\begin{acknowledgments}
The $\mu$SR measurements reported here were performed at the Swiss Muon Source, Paul Scherrer Institute, Switzerland. Parts of the work were supported by the Austrian FWF (Fonds zur F\"orderung der wissenschaftlichen Forschung) project P16370.
\end{acknowledgments}

\bibliography{amato_general}
\end{document}